\documentclass[twocolumn,english,aps,prd,reprint,floatfix,notitlepage,footinbib,preprintnumbers,superscriptaddress,longbibliography]{revtex4-1}
\pdfoutput=1
\usepackage{lmodern}

\usepackage[T1]{fontenc}
\usepackage[latin9]{inputenc}
\usepackage{geometry}
\usepackage{subfigure,lmodern, amsmath,amssymb, graphicx, pifont, adjustbox, bm, xcolor}
\usepackage{amsfonts}
\usepackage{enumitem}
\usepackage{comment}
\usepackage{mathtools}
\usepackage{float}
\usepackage{slashed}
\usepackage{ragged2e}
\usepackage{array}

\usepackage{nameref}

\usepackage{hhline}

\makeatletter\g@addto@macro\bfseries{\boldmath}\makeatother

\makeatletter\newcommand{\labeltext}[2]{%
  \def\@currentlabel{#1}%
  \label{#2}%
}
\makeatother

\usepackage{stackengine}
\usepackage{esint}
\usepackage[unicode=true,pdfusetitle,
 bookmarks=true,bookmarksnumbered=false,bookmarksopen=false,
 breaklinks=false,pdfborder={0 0 1},backref=false,colorlinks=true]
 {hyperref}
\hypersetup{
 pdfauthor={Clifford Cheung, Grant N. Remmen, Francesco Sciotti, Michele Tarquini},
 citecolor=black,linkcolor=black,urlcolor=black}

\newcommand{\appendixref}[1]{\hyperref[#1]{appendix~\ref{#1}}}
\def\equationautorefname~#1\null{eq.\,(#1)\null}
\usepackage{breakurl}
\usepackage{breakurl}
\usepackage[hang,flushmargin]{footmisc} 
\allowdisplaybreaks
\makeatletter

\usepackage{etoolbox}
\apptocmd{\thebibliography}{\justifying\setlength{\leftskip}{7.4mm}}{}{} 
 
 \usepackage{relsize}
\usepackage{babel}

\makeatletter
\def\simgt{\mathrel{\lower2.5pt\vbox{\lineskip=0pt\baselineskip=0pt
           \hbox{$>$}\hbox{$\sim$}}}}
\def\simlt{\mathrel{\lower2.5pt\vbox{\lineskip=0pt\baselineskip=0pt
           \hbox{$<$}\hbox{$\sim$}}}}
\makeatother

\usepackage{changepage}

\newcommand{\be}{\begin{equation}}
\newcommand{\ee}{\end{equation}}
\newcommand{\bea}{\begin{eqnarray}}
\newcommand{\eea}{\end{eqnarray}}
\newcommand{\Fig}[1]{Fig.~\ref{#1}}

\newcommand{\Eq}[1]{Eq.~(\ref{#1})}
\newcommand{\Eqs}[2]{Eqs.~(\ref{#1}) and (\ref{#2})}

\newcommand{\eq}[2]{\be\begin{aligned}#1 \label{#2}\end{aligned}\ee}
\newcommand{\mysec}[1]{\noindent {\bf #1.}---}

\newcolumntype{P}[1]{>{\centering\arraybackslash}p{#1}}

\usepackage{fix-cm}

\begin{document}

\preprint{CALT-TH 2025-027}

\title{Strings from Almost Nothing}

\author{Clifford Cheung}
\affiliation{Walter Burke Institute for Theoretical Physics, California Institute of Technology, Pasadena, CA 91125, USA}
\author{Grant N.~Remmen}
\affiliation{\scalebox{1}{Center for Cosmology and Particle Physics, Department of Physics, New York University, New York, NY 10003, USA}}    
\author{Francesco Sciotti}
\affiliation{IFAE and BIST, Universitat Aut\`onoma de Barcelona, 08193 Bellaterra, Barcelona, Spain}
\author{Michele Tarquini}
\affiliation{Walter Burke Institute for Theoretical Physics, California Institute of Technology, Pasadena, CA 91125, USA}

\begin{abstract}

\noindent We argue that string theory emerges inevitably from a few simple assumptions about physical scattering. Consistency alone requires that all tree-level four-point scattering amplitudes exhibit vanishing residues at prescribed values of the momentum transfer.  Assuming ultrasoft high-energy behavior, we then prove that the space of minimally consistent amplitudes, whose residues exhibit these mandated zeros and nothing more, collapses uniquely onto the celebrated Veneziano and Virasoro-Shapiro amplitudes of string theory. Similar logic also applies to five-point scattering.

\noindent 
\end{abstract}
\maketitle 

\mysec{Introduction}Are the laws of physics described by string theory?  
This question has elicited much controversy, despite being a well-posed scientific hypothesis that is either true or false.  The good news is that string theory has a sharp prediction: the emergence of very specific higher-spin excitations in gravitational scattering at or below the Planck scale.   The bad news is that experiment---the ultimate arbiter of all theories---will not offer any insight into this question in the foreseeable future, or perhaps ever.

In the absence of experimental input, it is tempting to commit faithfully either to string theory or its negation.  A less devoted approach is to ask to what extent string theory might follow from more modest or widely accepted assumptions.  One instead explores the space of consistent theories by ``bootstrapping'' all possible scattering amplitudes consistent with fundamental principles like unitarity, locality, and Lorentz invariance.  
Recent years have witnessed substantial progress in this area, for example on effective field theory constraints~\cite{Adams:2006sv, Nicolis:2009qm, Bellazzini:2014waa, Bellazzini:2015cra, Bellazzini:2019bzh, Camanho:2014apa, Arkani-Hamed:2020blm, Bellazzini:2020cot, Tolley:2020gtv,Bern:2021ppb, Chiang:2021ziz, Bellazzini:2021oaj, Karateev:2023mrb, Cheung:2025nhw, Arkani-Hamed:2021ajd, Cheung:2014ega, Remmen:2019cyz, Remmen:2020vts, Remmen:2020uze, Remmen:2022orj, Remmen:2024hry, Aoude:2024xpx, Cheung:2016yqr, Cheung:2016wjt, Green:2023ids, Freytsis:2022aho, Bellazzini:2019xts, Andriolo:2020lul, deRham:2017xox, Chandrasekaran:2018qmx, Jenkins:2006ia, Dvali:2012zc, Pham:1985cr, Ananthanarayan:1994hf, Pennington:1994kc}, the primal bootstrap \cite{Paulos:2017fhb, Cordova:2018uop, Guerrieri:2018uew, EliasMiro:2019kyf, Cordova:2019lot, Karateev:2019ymz, Guerrieri:2020kcs, Guerrieri:2020bto, CarrilloGonzalez:2022fwg, He:2021eqn, EliasMiro:2021nul, Guerrieri:2021tak, EliasMiro:2022xaa, Correia:2025uvc, deRham:2025vaq}, semidefinite programming and null constraints~\cite{Simmons-Duffin:2015qma, Caron-Huot:2020cmc, Caron-Huot:2021rmr, Caron-Huot:2022ugt, Albert:2022oes, Haring:2022sdp, Fernandez:2022kzi, Albert:2023jtd, Albert:2023seb, Li:2023qzs, Ma:2023vgc, Berman:2023jys, Beadle:2024hqg, Dong:2024omo, Berman:2024eid, Berman:2024kdh, Chang:2025cxc, Beadle:2025cdx, Berman:2025owb, Bellazzini:2025shd}, and bootstrapping string amplitudes and their deformations~\cite{Huang:2020nqy, Guerrieri:2021ivu, Cheung:2022mkw, Geiser:2022icl, Geiser:2022exp, Huang:2022mdb, Maldacena:2022ckr, Cheung:2023adk, bespoke, Arkani-Hamed:2023jwn, Bhardwaj:2024klc,  Jepsen:2023sia,  Gadde:2025fil,Figueroa:2022onw,Bhardwaj:2022lbz, Geiser:2023qqq, Zhiboedov, Caron-Huot:2016icg, Sever:2017ylk, Albert:2024yap, Cheung:2024uhn, Cheung:2024obl, Arkani-Hamed:2024nzc,  Jepsen:2025baw, Huang:2022mdb}.

In this paper, we argue for the uniqueness of string theory from the fewest assumptions possible.    All amplitude properties---the spectrum of masses and spins, high-energy behavior, and even its explicit mathematical form---are outputs of this bootstrap construction. 

\begin{figure}[t]
    \centering
    \includegraphics[width=0.625\columnwidth]{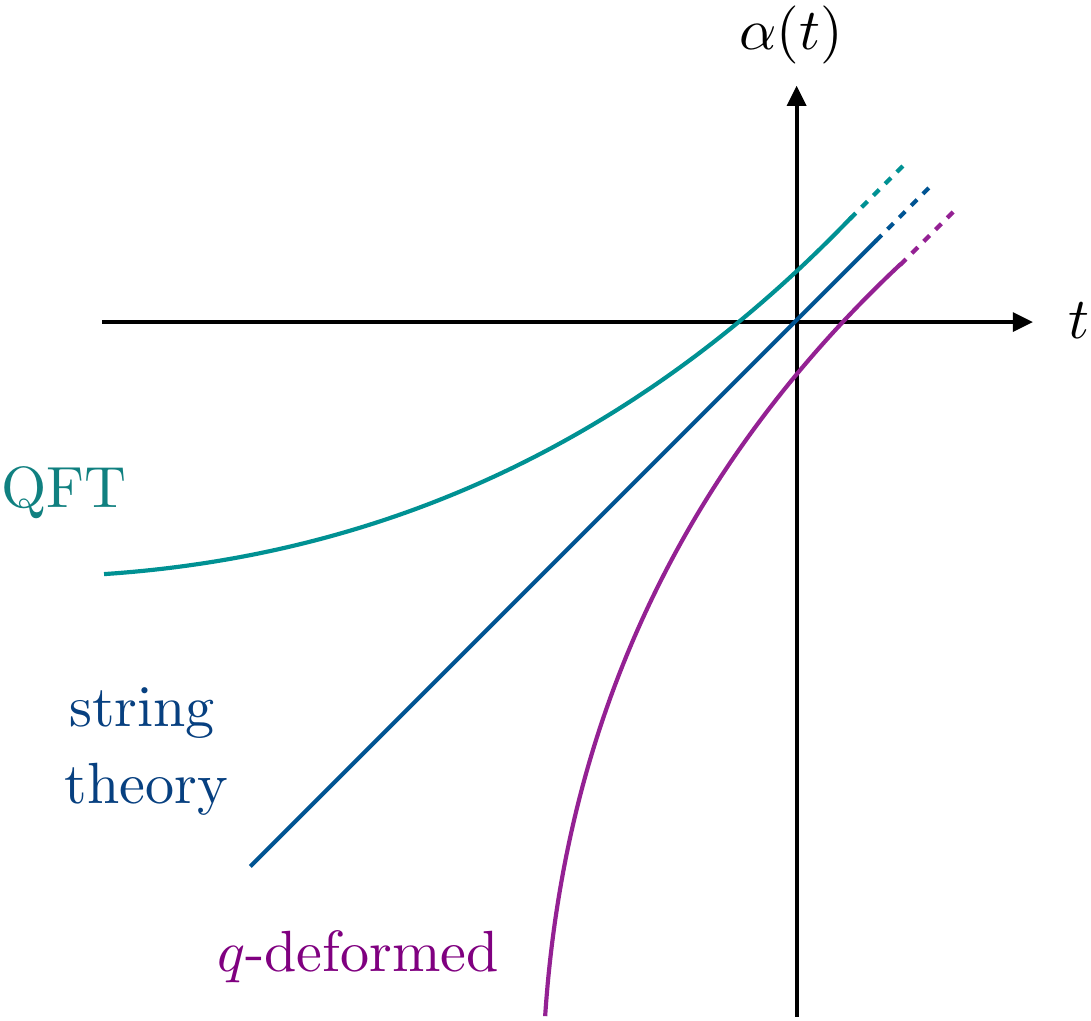}
    \caption{Schematic behavior of Regge trajectories for quantum field theory, string theory, and its $q$-deformations. 
    }
    \vspace{-1em}
    \label{fig:alpha_plot}
\end{figure}

At the heart of our analysis is a remarkable fact about {\it all tree-level, crossing-symmetric, Lorentz invariant, unitary amplitudes}.  
Without loss of generality, the spectrum of masses squared and spins is defined by unknown functions $\mu(n)$ and $J(n)$ indexed by an integer level $n\geq 0$. In the Regge limit, which corresponds to asymptotically large center-of-mass energy at fixed momentum transfer~\footnote{We work in mostly-plus metric signature throughout, where the Mandelstam invariants are $s = -(p_1+p_2)^2$, $t = -(p_2+p_3)^2$, and $u = -(p_1+p_3)^2$ for cyclically labeled incoming momenta.}, the amplitude scales as $(-s)^{\alpha(t)}$, where $\alpha(t)$ is the Regge trajectory. Here we will prove that the residue of the amplitude on any pole at large $s=\mu(n)$ must vanish when $\alpha(t)$ crosses certain negative integers.
These ``Regge zeros''  are mandated by consistency and apply to planar and nonplanar amplitudes at four-point and beyond for any $\mu(n)$, $J(n)$, and $\alpha(t)$.  
To single out string theory we further assume:

\vspace{-0.5em}

\begin{itemize}[
    labelindent=0pt,       % bullet sits exactly at the leftmargin
    leftmargin=0.2cm,        % inset the entire list 1cm from the normal LHS
    rightmargin=0.2cm,       % inset the entire list 1cm from the normal RHS
    labelsep=0.5em,        % space between bullet and text
    align=left             % make wrapped lines align under the first line
]
\item[] {\it i}) {\it Ultrasoftness.}\labeltext{{\it i})}{i}  The Regge trajectory $\alpha(t)$ is bijective for $t\,{<}\,0$ \footnote{Note that the $t\,{>}\,0$ behavior is unspecified, so a finite or infinite tower of spins is a priori allowed.} in such a way that the amplitude falls off faster than any power of the center-of-mass energy at fixed momentum transfer, so $\alpha(t) \,{\rightarrow}\,{-}\infty$ for $t\,{\rightarrow}\, {-}\infty$~\footnote{Strictly speaking, we impose this property on the real component of $\alpha(t)$, which controls the magnitude of the amplitude in the Regge limit, along the real $t$ axis. }.     As depicted in \Fig{fig:alpha_plot}, ultrasoftness is 
exhibited by string theory~\cite{Veneziano} but not its $q$-deformations~\cite{Coon:1969yw}, which diverge at finite $t$, nor  quantum field theory, whose softness is bounded~\cite{Martin,Buoninfante:2023dyd}. Ultrasoftness is equivalent to superpolynomial boundedness \cite{Zhiboedov} together with bijectivity.

\vspace{-0.2em}

\item[] {\it ii}) {\it Minimal Zeros.}\labeltext{{\it ii})}{ii}  Every zero of every residue is at the precise location of a Regge zero mandated by self-consistency.
Even though extra zeros are not inconsistent,  they are manifestly nonminimal because they go beyond what is required.

\end{itemize}

\noindent 
Given these modest conditions, our logic is simple: ultrasoft Regge behavior dictates an infinite set of points at which the residues of the amplitude are zero.  Bootstrapping the space of objects that exhibit these required zeros and nothing more, we land uniquely  on the Veneziano and Virasoro-Shapiro amplitudes of string theory.  

Note that our approach is quite distinct from that of Ref.~\cite{Caron-Huot:2016icg}, 
which argued for string uniqueness in tree-level four-point planar scattering while relying on a critical assumption about the  distribution of amplitude zeros.  Though conjectured to be universal, this property is now understood to be violated quite generally \cite{Zhiboedov}, leaving the question of string uniqueness an open one until now.

\medskip
\mysec{Regge Zeros}Consider the four-point amplitude $A(s,t)$ for identical color-ordered scalars interacting via tree-level exchanges.  This amplitude exhibits planar crossing symmetry, $A(s,t) = A(t,s)$, with simple poles~\footnote{Mathematically, we assume that the amplitude is a meromorphic function in the entire complex $s$ plane, including infinity.} at $s,t=\mu(n)$.
Without loss of generality we take $\mu(n)$ taken to be monotonically increasing.  The Regge limit of the amplitude is
\eq{
A(s,t) \sim f(t) (-s)^{\alpha(t)},
}{regge_planar}
which automatically encodes all discontinuities on the positive $s$ axis as a single branch cut \footnote{Recall that the famous Regge formula in \Eq{regge_planar} is an exact formula for asymptotically large $s$ everywhere in the complex $s$ plane except the positive real axis \cite{Sivers:1971ig}.  Case in point, any tree-level amplitude $A(s,t)$ is required to have simple poles on the real $s$ axis separated by regions of analyticity, which are all clearly absent from \Eq{regge_planar}.  In fact, Eq.~(\ref {regge_planar}) smears the
  discontinuities on the positive $s$ axis into a single branch cut, fusing the
  discrete spectrum of states into a continuum whose microscopic resolution is controlled by the gaps between poles.}.
Throughout this work, the symbol  $\sim$ will carry the precise meaning of equality up to subleading corrections in small $t/s$.  Occasionally we will toggle between large $s$ and large level $n$, in which case $\sim$ will indicate equality up to terms that fall off as a power of $1/n$ as $n$ goes to infinity.

The prefactor $f(t)$ is analytic except for simple poles at $t=\mu(n)$, whose residues are  polynomials in $s$ of degree $J(n)$, the maximal spin at level $n$~\footnote{The case of infinite $J(n)$ at finite $n$ corresponds to an infinite tower of spins residing on an isolated,  narrow resonance.  This describes a propagating mode of unbounded spatial extent, in violation of locality~\cite{Caron-Huot:2020cmc,Bern:2021ppb,Huang:2022mdb}.}.  Hence, the spectrum and the leading Regge trajectory satisfy 
\eq{
 \alpha(\mu(n))  = J(n)  \textrm{ for integer }  n \geq 0,
}{mass_spin_relation}
which is also applicable without any loss of  generality.

\begin{figure*}[t]
    \centering
    \includegraphics[width=18cm]{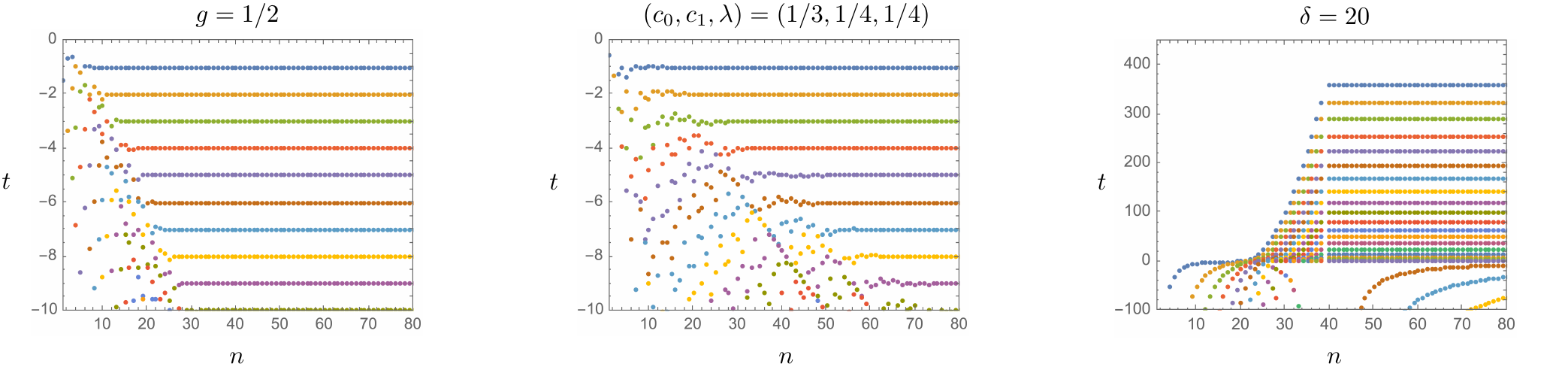} \vspace{-0.75cm}
    \caption{The residues of tree-level four-point amplitudes exhibit Regge zeros at asymptotically large level $n$.  Shown here are the real components of the zeros in $t$ for the residues defined in Eqs.~(\ref{R_Gross}), (\ref{R_HZ}), and (\ref{R_bespoke}), for levels $n=1 \textrm{ to } 80$. 
    }
        \vspace{-1em}
    \label{fig:zero_plot}
\end{figure*}

We are now ready to derive a surprising and universal fact about all tree-level, color-ordered, four-point amplitudes.
To begin, we observe that when $\alpha(t)$ is an integer, the Regge formula $(-s)^{\alpha(t)}$ sheds its branch cut and produces a strict power law.  This function is obviously analytic in $s$ in the neighborhood of complex infinity---with the possible exception of the region exactly on the positive real axis, which is the only place the Regge formula need not hold.  Thus we are guaranteed that the amplitude is analytic on the open ball surrounding the north pole of the stereographic sphere, minus the north pole itself and the positive real axis.  However, since a power law function is automatically analytic on the positive real axis, the uniqueness of analytic continuation implies that the amplitude cannot have nonanalyticities there either.  This forbids the possibility of an accumulating series of poles on the approach to infinity, and thus the tower of resonances at asymptotically large $n$ must have {\it vanishing} residues for integral $\alpha(t)$.

Another way to see this property is to directly compute the contour integral of $A(s,t)$ along a circle of large radius $s_*$ in two ways: by explicit computation \footnote{Here we directly integrate \Eq{regge_planar}, which exhibits a branch cut along the positive real $s$ axis corresponding to all  physical discontinuities.   To compute \Eq{arc_integral_planar} we transform to polar coordinates, $s=s_* e^{i\theta}$, and integrate around the branch cut in the domain $0 < \theta <2\pi $.
},
\eq{
\oint^{s_*} ds\, A(s,t) 
 \,{\sim}\, f(t) s_*^{\alpha(t) + 1}  \frac{\sin \pi \alpha(t)}{\alpha(t)+1} \,{\sim}\, \sum^{n_*}_n  R(n,t),
}{arc_integral_planar}
and by contour deformation to obtain the sum over enclosed residues, $R(n,t) = \lim_{s\rightarrow\mu(n)}(\mu(n)\,{-}\,s)A(s,t)$, up to the largest integer $n_*$  bounded by  $s_* > \mu(n_*)$.  Since $(-s)^{\alpha(t)}$ is a monomial,  \Eq{arc_integral_planar} equals one for $\alpha(t)\,{=}\,{-}1$ and zero for $\alpha(t) \,{\neq}\, {-}1$.   Either way, \Eq{arc_integral_planar} is {\it independent} of $n_*$, so the residues at large level are zero~\footnote{This condition is trivially satisfied when $\alpha(t)$ is a constant integer with no $t$ dependence whatsoever. In this case the amplitude is a rational function, so for sufficiently large $s_*$ there are no poles nearby, and thus any residue computed there will be zero. While our logic still applies, this scenario is obviously uninteresting, so we will assume throughout that the leading Regge trajectory $\alpha(t)$ actually depends on $t$. }.

To extract the asymptotic behavior of the residue we subtract a pair of concentric arc integrals to obtain
\eq{
R(n,t) \sim  f(t) \mu'(n)\mu(n)^{\alpha(t) } \sin \pi \alpha(t),
}{residue_form_planar}
where $\mu'(n) = \mu(n) -\mu(n-1)$ is shorthand for a finite difference and we assume that $\mu'(n) \ll \mu(n)$.
Note again that \Eq{residue_form_planar} is valid in the Regge limit, so it is an exact equality up to subleading corrections at large $n$.

The $ \sin \pi \alpha(t)$  in \Eq{residue_form_planar} implies that $R(n,t)=0$ when $\alpha(t)$ is an integer, subject to two caveats.  First, since $R(n,t)$ is nonsingular and $f(t)$ has poles at $t\,{=}\,\mu(n')$ for any integer $n'\,{\geq}\,0 $, those poles must cancel exactly against corresponding zeros in  $ \sin \pi \alpha(t)$.  Thus, a genuine zero of \Eq{residue_form_planar} requires the off-resonance condition $t\,{\neq}\,\mu(n')$ for all $n'\,{\geq}\, 0$.
Second, unitarity implies that $R(n,t)\,{=}\,\sum_{\ell{=}0}^\infty a_{n,\ell} G_\ell^{(D)}(1\,{+}\,2t/\mu(n))$ with $a_{n,\ell}\,{\geq}\, 0$.  For $\ell> 0$, the $D$-dimensional Gegenbauer polynomials satisfy $G_\ell^{(D)}(x)\,\,{>}\,\, G_\ell^{(D)}(1) \,\,{>}\,\,0$  when $x>1$.  Since $\mu(n),\,J(n)\,\,{>}\,\,0$ at large $n$, we see that $R(n,t) > R(n,0)>0$ for $t\,{>}\,0$, so this region has no residue zeros.

 In summary, \Eq{residue_form_planar}  implies that the residue $R(n,t)$ at large $n$, corresponding to the Regge limit, must vanish on the support of Regge zeros,
\eq{
\alpha(t)+r =0  ,
}{zero_locus_planar}
for any integer $r$ such that the real roots of \Eq{zero_locus_planar}  satisfy $t<0$ and $t \neq \mu(n')$ for any integer $n'\geq 0$.  Here \Eq{zero_locus_planar} is the exact location of the zeros of $R(n,t)$ in the strict Regge limit defined by infinite $n$.  Conversely, at finite $n$, the residue $R(n,t)$ will by continuity have zeros at values of $t$ that are shifted from the solutions of \Eq{zero_locus_planar} by corrections that decouple as $n$ grows.  Regge zeros are required by consistency in any  unitary, tree-level, color-ordered, four-point amplitude.   
Since \Eq{zero_locus_planar} holds in the Regge limit,  any Regge zero of  $R(n,t)$ will also be a Regge zero of $R(n',t)$ for $n'>n$ since the approximation of the Regge limit only improves as $n$ grows.

Softer Regge behavior implies more Regge zeros.  For example, the Veneziano amplitude of string theory scales as $ \Gamma(-s) \Gamma(-t)/\Gamma({-}s\,{-}\,t) \sim s^t$,
with unboundedly fast falloff for sufficiently negative $t$.  The expected Regge zeros at $t+r=0$ appear in the string residues,
\eq{
R_{\rm str}(n,t) &= \frac{1}{n!} \prod_{r=1}^n (t+r) 
\sim n^t.
}{}
Regge zeros are ubiquitous~\footnote{Note that the hypergeometric~\cite{Cheung:2023adk} and ${}_4 F_3$ satellite~\cite{Albert:2024yap} amplitudes exhibit power-law Regge behavior reminiscent of quantum field theory, where $\alpha(t)$ crosses few if any integers, yielding few if any Regge zeros, as expected.}, but they are often difficult to ascertain from analytic expressions.  Consider three examples: the residues of the simplest deformed worldsheet amplitude in the family of Ref.~\cite{Gross},
$\int_0^1 u^{-s-1}(1-u)^{-t-1} e^{g u(1-u)} du$~\footnote{
In the notation of Ref.~\cite{Gross}, this deformed worldsheet integral corresponds to taking $\tilde f_4(u) = g/4$, a constant, and is hence the simplest possible model in that family. We can write the amplitude in closed form,\vspace{-1mm}
\begin{equation*}
\hspace{6mm} A(s,t) = \frac{\Gamma(-s)\Gamma(-t)}{\Gamma(-s-t)}{}_2 F_2 \left[\begin{array}{c}-s,-t \\ -\frac{s+t}{2},\frac{1-s-t}{2}\end{array};\tfrac{g}{4}\right],\vspace{-1mm}
\end{equation*}
by expanding the exponential and resumming.}, 
  \eq{
 R_{\rm str}(n,t) \,{}_2 F_2 \left[\begin{array}{c}{\scriptstyle-n},{\scriptstyle-t} \\ -\frac{n+t}{2},\frac{1-n-t}{2}\end{array};\tfrac{g}{4}\right],
}{R_Gross}
the three-parameter amplitudes explored in Ref.~\cite{Zhiboedov}, 
\eq{
R_{\rm str}(n,t)\, {}_3 F_2\left[\begin{array}{c} {\scriptstyle-n},{\scriptstyle-t},{\scriptstyle-c_0-c_1(n+t)} \\ -\frac{n+t}{2},\frac{1-n-t}{2}\end{array};\lambda \right],
}{R_HZ}
and one of the bespoke amplitudes of Ref.~\cite{bespoke},
\eq{
2(n+\delta)(R_{\rm str}(n, -\delta + \sqrt{t}) + R_{\rm str}(n, -\delta - \sqrt{t})).
}{R_bespoke}
Here $ R_{\rm str}(n,t)$ appears in all of these residues, but its zeros are either cancelled by poles or scrambled by changes of arguments.  Nevertheless, Eqs.~(\ref{R_Gross}), (\ref{R_HZ}), and (\ref{R_bespoke}) are all polynomials in $t$ that exhibit  their expected Regge zeros at large $n$, as emerges quite vividly via direct numerical evaluation in   \Fig{fig:zero_plot}.   Since \Eqs{R_Gross}{R_HZ} have the same spectrum and Regge behavior as the Veneziano amplitude, they also have Regge zeros that asymptote to $t+r=0$.    In contrast, \Eq{R_bespoke}  describes a modified spectrum, $\mu(n) = (\delta+n)^2$, and Regge behavior,
$A(s,t)\sim (\sqrt{s})^{-\delta+\sqrt{t}}+(\sqrt{s})^{-\delta -\sqrt{t}}$, so its Regge zeros are at $t=(\delta-r)^2$ for $t$ off-resonance, as expected.

Equation~\eqref{residue_form_planar} implies a quite constraining residue scaling, $R(n,t)\,\,{\sim}\,\, \mu(n)^{\alpha(t)}$.  In fact, this property rules out any amplitude with a single Regge trajectory.  By definition, the residues of any such amplitude describe a single state of spin $n$ state at level $n$, so $R(n,t)\,{=}\, c(n)\, G_n^{(D)}\!\!\left(1\,{+}\,2t/\mu(n)\right)$. 
This implies that  $\partial_t^k \log R(n,t) |_{t=0} \sim (n^2/\mu(n))^k$ at large $n$, modulo multiplicative factors. 
However, from the residue scaling we know that $\partial_t^k \log R(n,t) |_{t=0}\sim 
 \log \mu(n)$,
indicating the same scaling with $n$ for all $k$, which is a contradiction.  Similar conclusions were obtained via different methods in Refs.~\cite{Eckner:2024ggx, Eckner:2024pqt}. All of these no-go results assume that the single Regge trajectory extends to infinity without any additional high-energy states.

\medskip
\mysec{Uniqueness Argument}We are now equipped to derive a uniqueness principle for string theory, assuming \ref{i}~ultrasoft Regge behavior, and \ref{ii}~residue zeros comprised only of Regge zeros required by consistency.   The former imposes an infinite number of Regge zeros at asymptotically large level $n$.  The latter forbids any extra zeros that are not Regge zeros, even at finite $n$.  As noted earlier, Regge zeros necessarily accumulate for increasing $n$, so  \ref{ii}~implies a nested structure whereby every zero of $R(n,t)$ is also a zero of $R(n',t)$ for $n'>n$.

Taking the ratio of \Eq{residue_form_planar} evaluated at adjacent levels, followed by a logarithm, we obtain
\eq{
\alpha(t) \log\frac{\mu(n)}{\mu(n{-}1)} + \log\frac{\mu'(n)}{\mu'(n{-}1)} \sim \log\frac{R(n,t)}{R(n{-}1,t)}.
}{alpha_sol_planar}
Up to a normalization and shift, \Eq{alpha_sol_planar} implies the Regge trajectory is equal to $ \log P(n,t)$, where $P(n,t) = R(n,t)/R(n\,{-}\,1,t)$ is a rational function of $t$.  Assumption~\ref{ii} says that the residues $R(n,t)$ and $R(n-1,t)$ only exhibit Regge zeros, and we argued earlier that any Regge zeros of $R(n-1,t)$ are also Regge zeros of $R(n,t)$, thus canceling in the ratio and leaving $P(n,t)$ to be a polynomial in $t$.

At asymptotically large level $n$, let us define the degree of this polynomial to be $k\geq0$, so  $P(n,t)\sim t^k$ for sufficiently large $t$.  For $k>0$, this implies a logarithmic Regge trajectory $\alpha(t) \propto k \log t$, where  $\propto$ denotes equality up to constant shifts and rescalings. Plugging into \Eq{mass_spin_relation}, we learn that $J(n) \propto k \log \mu(n) $.  For $k=1$, this is precisely the $q$-deformed spectrum of the Coon amplitude~\cite{Coon:1969yw}, which has $P(n,t) \sim q+q(q{-}1)t$ \footnote{The original Coon amplitude~\cite{Coon:1969yw}, whose residues are consistent with unitarity only for $q<1$, exhibits an accumulation point in the spectrum and is neither dual resonant nor meromorphic, on account of a branch cut dressing factor~\cite{Cheung:2023adk,Coon:1972qz}.
While amplitudes of this form are realized in nonperturbative string theoretic constructions~\cite{Maldacena:2022ckr}, they depart from our assumption of tree-level dynamics, so we leave them to future work.}.  More generally, since $k>0$ we see that $\alpha(t)\rightarrow \infty$ as $t\rightarrow -\infty$, so these logarithmic trajectories violate our assumption~\ref{i}.  We will not consider them further.

For $k=0$,  the terms in $P(n,t)$ that are subleading at large $n$ yield the leading $t$-dependence of the Regge trajectory.  This happens in the Veneziano amplitude, where $P(n,t)\sim 1+t/n$.  As the subleading terms are polynomial, $\alpha(t)$ is as well.    We focus on this case next.

All residues exhibit the Regge zeros in \Eq{zero_locus_planar} for asymptotically large $n$, but assumption~\ref{ii} enforces the stronger condition that these Regge zeros comprise {\it the exact locations of all zeros} of the residue for all $n$ and $t$, so $R(n,t)$ is a product of factors of $\alpha(t)+r$.  Since unitarity restricts all Regge zeros to the region $t\,{<}\,0$,   assumption~\ref{i} implies that $\alpha(t)\,{<}\,\alpha(0)$, which on the Regge zero requires $r \,{=}\, {-}\alpha(t) \,{>}\, {-}\alpha(0)$.  Here  we assume a minimal value of $r=1$, since any finite offset will just shift $\alpha(t)$ in a way that leaves our conclusions unchanged.  Concretely, the residue takes the form \footnote{Here we have implicitly assumed that {\it every} zero of $\alpha(t)+r=0$ appears in $R(n,t)$ as a block.  Said another way, $R(n,t)$ does not include partial branches of solutions where only a subset of the zeros of $\alpha(t)+r=0$ appear.  This structure is justified because of the underlying logic of the Regge zeros: if one solution of $\alpha(t)+r=0$ appears in $R(n,t)$, then $n$ is sufficiently large to justify the arc integral argument that implies the Regge zeros, and thus the other solutions of $\alpha(t)+r=0$ should be exhibited as well.}
\eq{
R(n,t) &= c(n) \prod_{r=1}^{N(n)} (\alpha(t)+r)^h ,
 }{residue_ansatz_polynomial_planar}
where $h$ is a constant multiplicity of each Regge zero, $c(n)$ is an arbitrary normalization, and we extend the range of $r$ up to $N(n) = J(n)/h\,{\rm deg}(\alpha)$.
From \Eq{residue_ansatz_polynomial_planar} we see that $R(n,t) \propto R_{\rm str}(N(n),\alpha(t))^h  \sim N(n)^{h\alpha(t)}$, while \Eq{residue_form_planar} implies that $R(n,t)\sim \mu(n)^{\alpha(t)}$. Comparing these representations, we learn that 
$\mu(n) \sim N(n)^h \propto J(n)^h$. 
Plugging into \Eq{mass_spin_relation}, we see that $\alpha(J(n)^h) \sim \alpha(\mu(n)) = J(n)$.  Using our earlier argument that $\alpha(t)$ is polynomial, we conclude that $h=1$ and so $\alpha(t) \propto t$.
This establishes a linear spectrum and Regge trajectory,
\eq{
\mu(n)   \propto J(n)  \quad \textrm{and}\quad \alpha(t) \propto   t,
}{linear_everything}
which is one of the main results of this paper.  In conclusion, our assumptions  \ref{i} and  \ref{ii} imply the Chew-Frautschi scaling and linear Regge behavior that are the defining properties and historical raison d'\^{e}tre~\cite{Regge,Chew:1961ev,Chew:1961yz,Veneziano} of string theory.  Note that it is straightforward to also derive  that $J(n)= n$, which we detail in App.~\ref{sec:spin_linear}.

With $J(n)= n$ and $\alpha(t)\propto t$, the residue ansatz in \Eq{residue_ansatz_polynomial_planar} precisely reduces to the ``level truncation'' ansatz proposed in Refs.~\cite{Cheung:2024uhn, Cheung:2024obl}, which is solved analytically to obtain a three-parameter space of amplitudes.
Within this family, the only object that exhibits arbitrarily soft Regge behavior is $\Gamma(-s)\Gamma(-t)/\Gamma({-}s\,{-}\,t)$, which is precisely the Veneziano amplitude~\footnote{To be precise, the result of this bootstrap procedure is the $Z$-theory amplitude~\cite{Carrasco:2016ldy} for colored external scalars, which is trivially related to the amplitudes of the superstring and bosonic string by a simple prefactor and shift of the Mandelstam invariants, respectively.  The latter correspond to offsets in $\alpha(t)$,  $J(n)$, or $\mu(n)$, which can also be used to generate so-called satellite amplitudes.}.  Thus, our assumptions~\ref{i} and \ref{ii} actually uniquely bootstrap the Veneziano amplitude in its full mathematical form!

Assumption~\ref{ii} is our least conservative condition, but it can be relaxed by including extraneous zeros in the residue in  \Eq{residue_ansatz_polynomial_planar}.  For example, if \Eq{residue_ansatz_polynomial_planar} is augmented by a level-independent factor $g(t)$, as might occur for external states with spin, all of our arguments still apply.  For a level-dependent factor, $g(n,t)$, our logic persists if the number of extraneous zeros is parametrically smaller than the number of Regge zeros.

\medskip

\mysec{Beyond Planar}Up until now we have focused on planar amplitudes.  However, our logic generalizes straightforwardly to the case of nonplanar scattering.
Consider the four-point amplitude $A(s,t)$ for identical massless scalars interacting via tree-level exchanges.  The amplitude is crossing symmetric under permutations of $s,t,u$ and has poles at $s,t,u=\mu(n)$.   The Regge behavior at fixed $t$ is parameterized by \vspace{-1mm}
\eq{
A(s,t) \sim f(t)(-s)^{\frac{\alpha(t)}{2}}(-u)^{\frac{\alpha(t)}{2}},
}{}
which is crossing symmetric  under the exchange of $s$ and 
$u=-s-t$. 
As before, we compute the big arc integral along a circle of radius $s_*$ in the $s$ plane, \vspace{-1mm}
\eq{
\oint^{s_*}d(su)  \, A(s,t) 
& \sim  f(t)s_*^{\alpha(t)+2}  \frac{\sin\tfrac{1}{2} \pi \alpha(t)}{\alpha(t)+2},
}{}
and deform the contour to equate it to a weighted sum of residues, $\sum^{n_*}_n  (\mu(n)+t/2) R(n,t)$.
Again taking the difference of concentric arc integrals, we obtain the residue
\eq{
R(n,t) \sim  f(t) \mu'(n)\mu(n)^{\alpha(t) }  \sin\tfrac{1}{2} \pi \alpha(t),
 }{residue_form_nonplanar}
implying  the  Regge zeros,
 \eq{
\alpha(t) +2r=0 ,
 }{} 
for integer $r$, again with the caveat that any real roots must satisfy $t\,{<}\,0$ and $t\neq \mu(n')$ for any integer $n' \geq 0$.
 
Equation~\eqref{residue_form_nonplanar} is nearly identical to \Eq{residue_form_planar}, so applying the same logic as in the planar case, we deduce that $\alpha(t)$ is a polynomial.
As before, assumption~\ref{ii} says that the residue is composed entirely of Regge zeros, so \vspace{-1mm}
 \eq{
R(n,t) &= c(n) \prod_{r=1}^{N(n)} (\alpha(t)+2r)^h .
 }{R_prod_VS}
Since  $R(n,t) \propto R_{\rm str}(N(n),\alpha(t)/2)^h  \sim N(n)^{h\alpha(t)/2}$ and the asymptotic behavior is $R(n,t) \sim \mu(n)^{\alpha(t)}$, we find that $\mu(n) \sim N(n)^{h/2} \propto J(n)^{h/2}$.   Together with \Eq{mass_spin_relation}, this implies that
$\alpha(J(n)^{h/2}) \sim \alpha(\mu(n))=J(n)$.  Since $\alpha(t)$ is a bijective polynomial, we deduce that $h\,{=}\,2$, so $\alpha(t)\propto t$ and thus $\mu(n)\propto J(n)$.  Thus the nonplanar amplitude also has the linear spectrum and Regge trajectory defined in \Eq{linear_everything}.  As in the planar case, App.~\ref{sec:spin_linear} implies $J(n)= n$, so our nonplanar uniqueness arguments lead to the level truncation ansatz of Ref.~\cite{Cheung:2024obl}, whose only solution with ultrasoft Regge behavior is $\Gamma(-s)\Gamma(-t)\Gamma(-u)/\Gamma(1+s)\Gamma(1+t)\Gamma(1+u)$, which is the Virasoro-Shapiro amplitude~\cite{Virasoro,Shapiro}.

\medskip
\mysec{Beyond Four-Point}The Regge behavior of five-point string amplitudes also mandates Regge zeros in their corresponding residues.  In fact, these zeros are sufficient to uniquely bootstrap the five-point amplitude.

Here we study a worldsheet integral representation of the five-point string amplitude constructed in Ref.~\cite{Arkani-Hamed:2024nzc}, \vspace{-1mm}
\eq{
&A_{\rm str,5} = \int_0^1 \frac{du_{13}}{u_{13}(1{-}u_{13})}\frac{du_{14}}{u_{14}(1{-}u_{14})}F(u_{13},u_{14}),\\
&F(u_{13},u_{14}) = \frac{u_{13}^{X_{13}}u_{14}^{X_{14}}(1{-}u_{13})^{X_{24}}(1{-}u_{14})^{X_{35}}}{(1{-}u_{13}u_{14})^{X_{24}{+}X_{35}{-}X_{25}}},
}{A5}
where $X_{ij} = (p_i + \cdots + p_{j-1})^2$~\footnote{This representation of the string amplitude is obtained by transforming the Koba-Nielsen integral into one involving the $u$-variables~\cite{Koba:1969rw,BardakciRuegg,ChanTsou,Gross,BinGeom,Arkani-Hamed:2019mrd,Arkani-Hamed:2023jwn,Zeros,Splits}, with integrand going like $u_{13}^{X_{13}}u_{14}^{X_{14}}u_{24}^{X_{24}} u_{25}^{X_{25}}u_{35}^{X_{35}}$.
One then uses the $u$-equations that relate the different $u$-variables among each other to eliminate all but two of them~\cite{Arkani-Hamed:2024nzc}.}.  
In the Regge limit $X_{13},X_{14}\,{\rightarrow}\,\infty$,  the factors of $u_{13}^{X_{13}}$ and $u_{14}^{X_{14}}$ localize the integral to the saddle at $u_{13} = 1\,{-}\,X_{13}^{-1}$ and $u_{14} = 1 \,{-}\, X_{14}^{-1}$, with scaling $F(1{-}X_{13}^{-1},1{-}X_{14}^{-1}) \sim X_{13}^{X_{35}{-}X_{25}} X_{14}^{X_{24}{-}X_{25}}  (X_{13}{+}X_{14})^{X_{25}{-}X_{24}{-}X_{35}}$. 
Since we have $F(1-\epsilon,1-\epsilon)\sim \epsilon^{X_{25}}$, the integral converges provided $X_{25}>0$, just like the ${}_3 F_2$ closed-form expression for the five-point string amplitude~\cite{Bialas:1969jz}, which can be analytically extended to the entire kinematic space using the formulas of Ref.~\cite{Arkani-Hamed:2024nzc}.
In terms of  the large $s$-like invariants, $s_1=-X_{13}$ and $s_2=-X_{14}$, and the fixed $t$-like invariants, $t_{23} = -X_{24}$,
$t_{34}= -X_{35}$, and 
$t_{51} = -X_{25} $, we obtain our final result for the double Regge limit of the five-point string amplitude when $s_1,s_2\rightarrow\infty$,\vspace{-1mm}
 \eq{
\hspace{-2mm} A_{\rm str,5}  \sim  f(a_1{,}a_2{,}a_{12})  ({-}s_1)^{a_1} \! ({-}s_2)^{a_2}  ({-}s_1{-}s_2)^{a_{12}}\! ,\hspace{-2mm}
}{Regge_5pt}
where we have defined $a_1= t_{51} -t_{34}$, $a_2= t_{51} -t_{23}$, and $a_{12} =t_{23} + t_{34} - t_{51}$.
In analogy with our analysis at four-point, we now compute the double big arc integral, $\oint^{s_{1*}}\oint^{s_{2*}} ds_1ds_2\, A_{\rm str,5}$, which, by deforming both arc integral contours into a double sum over residues, equals $\sum^{n_{1*}}_{n_1}  \sum^{n_{2*}}_{n_2}  R_{\rm str,5}(n_{1},n_{2}, t_{23}, t_{34}, t_{51})$. Inserting \Eq{Regge_5pt}, we compute the double big arc integral to be $f$ times
\eq{
\hspace{-2mm} 
\sum_{b=0}^{a_{12} }\! s_{1*}^{a_1{+}1{+}b} \! s_{2*}^{a_2{+}1{+}a_{12}{-}b} 
\left(\substack{a_{12}\\ b}\right)
\frac{\sin \pi a_1\sin\pi(a_2{+}a_{12})}{(a_1\!{+}1\!{+}b)(a_2{+}1\!{+} a_{12}\!{-} b)} ,\hspace{-2mm}
}{}
where have taken $a_{12} \geq0$ to be a nonnegative integer so that the binomial expansion is finite.  Computing the finite differences in $n_{1*}$ and $n_{2*}$, we obtain\vspace{-1mm}
\eq{
&\frac{R_{\rm str,5}(n_{1},n_{2}, t_{23}, t_{34}, t_{51})}{\mu'(n_{1})\mu'(n_{2})\mu(n_1)^{a_1}\mu(n_2)^{a_2{+}a_{12}}} \\& \quad \sim f \sum_{b=0}^{a_{12} } \left(\tfrac{\mu(n_{1})}{\mu(n_2)}\right)^{b} 
\left(\substack{a_{12}\\ b}\right)
\sin \pi a_1\sin\pi(a_2{+}a_{12}),
}{residue_form_5pt}
which stringently constrains the asymptotics and zeros of the five-point residues.  By inspection, from \Eq{residue_form_5pt} we see that the residue vanishes when $a_1$, $a_2$, and $a_{12}$ are integers.   Recalling that $X_{25}>0$ for convergence, we deduce the locus of the Regge zeros for five-point residues at large level,
\eq{
  -(a_1 +a_2 + a_{12} )\geq 1 \quad  \textrm{and} \quad a_{12}\geq 0 ,
}{zeros_5pt}
for integers $a_1,a_2,a_{12}$.

It is straightforward to verify the presence of these Regge zeros in the known expression for the five-point string residue $R_{\rm str,5}(n_1,n_2,t_{23}, t_{34}, t_{51})$ \cite{Arkani-Hamed:2024nzc},\vspace{-1.5mm}
\begin{equation}
\hspace{-1.5mm}\frac{ (1{+}t_{23})_{n_1} (1{+}t_{34})_{n_2} }{n_1! n_2!} {}_3 F_2\left[\!\begin{array}{c}{\scriptstyle -n_1},{\scriptstyle -n_2},{\scriptstyle {-}t_{23}{-}t_{34}{+}t_{51}} \\ {\scriptstyle-n_1-t_{23}},{\scriptstyle-n_2-t_{34}}\end{array};1\right].\hspace{-0.5mm}
\label{eq:5ptres}
\end{equation}
\vspace{-3.5mm}

\noindent We see that there are indeed zeros for $a_1+r_1=a_2+r_2=a_{12}+r_{12}-r_1-r_2=0$, where $|r_1|\leq n_1$, $|r_2|\leq n_2$, and $1\leq r_{12}\leq r_1+r_2$.  This agrees exactly with \Eq{zeros_5pt} at large $n_1$ and $n_2$.  These Regge zeros can be used to constrain a five-point residue ansatz $R(n_1,n_2,t_{23}, t_{34}, t_{51})=\sum_{i=0}^{n_1} \sum_{j=0}^{n_2} \sum_{k=0}^{{\rm min}(n_1-i,n_2-j)} \lambda_{i,j,k}(n_1,n_2) t_{23}^i t_{34}^j t_{51}^k$ that encodes all possible exchanges of particles with maximum spins $n_{1,2}$ on the poles at $s_{1,2} = n_{1,2}$. Enforcing the Regge zeros fixes all $\lambda_{i,j,k}(n_1,n_2)$ to values corresponding {\it precisely} to the known residues of the string in \Eq{eq:5ptres}, modulo a normalization $c(n_1,n_2)$.

Since Regge zeros fix the five-point string residue, it is natural to ask about the structure of its zeros more generally, which we describe in detail in App.~\ref{app:5pt_level_truncation}.
Setting all three $t$-like Mandelstams to certain negative integers, one finds that all but a finite number of the residues vanish, reducing an infinite dual resonant sum over all levels $(n_1,n_2)$ down to a finite subset, making the amplitude rational in $(s_1,s_2)$.
As shown in Ref.~\cite{Cheung:2024uhn}, four-point crossing symmetry sculpts out an analytically solvable space of amplitudes that exhibit level truncation.  Analogously, in five-point scattering the  full cyclic invariance of the external legs produces a similarly solvable algebraic problem.  As detailed in App.~\ref{sec:5pt_normalizations}, demanding cyclic invariance in fact fixes all $c(n_1,n_2)=1$.
This implies that our bootstrapped amplitude is uniquely fixed by its residue zeros to match the five-point string amplitude.  The $q$-deformed version of our planar analyses is presented in App.~\ref{sec:qdef}. 

\medskip

\mysec{Discussion}We have demonstrated a universal link between the Regge behavior of an amplitude and the zeros of its residues.  This result applies to any four-point, tree-level, crossing-symmetric, Lorentz invariant, unitary amplitude.   Amplitudes with softer Regge behavior are constrained by more Regge zeros. Assuming a minimality condition that every zero of a residue is a Regge zero mandated by consistency, we proved that any ultrasoft amplitude has a stringy spectrum of masses squared that is linear in spin, $\mu(n)\propto  J(n)= n$, together with a linear Regge trajectory, $\alpha(t)\propto t$.  This reproduces the level truncation ansatz of Ref.~\cite{Cheung:2024uhn}, from which the exact mathematical formula for the Veneziano amplitude is derived uniquely.  While extra zeros are certainly allowed, the notion of minimal zeros gives an operational definition of the ``simplest consistent amplitude,'' which is the Veneziano amplitude itself.
Last but not least, we established analogous Regge zero and uniqueness properties beyond the planar four-point limit, successfully bootstrapping the four-point closed-string and five-point open-string amplitudes.

\medskip
\bigskip

\noindent {\it Acknowledgments:} 
We thank Justin Berman, Simon Caron-Huot, Henriette Elvang, Aaron Hillman, and Sasha Zhiboedov useful discussions.
C.C.~and M.T.~are supported by the Department of Energy (Grant No.~DE-SC0011632) and by the Walter Burke Institute for Theoretical Physics. 
G.N.R. is supported by the James Arthur Postdoctoral Fellowship at New York University. F.S.~is
supported by the research grants 2021-SGR-00649, PID2023-146686NB-C31, and funding from the European Union NextGenerationEU(PRTR-C17.I1). F.S.~also thanks the Walter Burke Institute for Theoretical Physics for its hospitality during the completion of this work.

\bibliographystyle{utphys-modified}
\bibliography{uniqueness}

\bigskip

\appendix

\setcounter{equation}{0}
\renewcommand{\theequation}{A\arabic{equation}}

%%%%%%%%%%%%%%%%%%%%%%%%%%%%%%%%%

\mysec{Appendix A: Linear Spin Spectrum\labeltext{A}{sec:spin_linear}}Under our stated assumptions, we have shown that $\mu(n)\sim J(n)$.  Here we will furthermore deduce that $J(n)= n$.  
To begin, we define $t_*$ to be the locus of the Regge zero at $\alpha(t_*)+r_*=0$, where $r_* = J(n_*)$ is the maximum spin of the residue at level $n_*$.  
We compute the big arc integral,
\eq{
\oint^{s_*} ds \,s^{r-1} A(s,t_*)  \sim  s_*^{\alpha(t_*)+r} =0 ,
}{}
which is zero for any $r$ in the range $1 \leq  r <r_*$.  Hence the number of constraints scales as $r_* = J(n_*)$ at large level.  
We can also rewrite the big arc integral as a sum of residues in the usual way,
\eq{
\oint^{s_*} ds \,s^{r-1} A(s,t_*) \sim \sum_{n =0}^{n_*-1} \mu(n)^{r-1} R(n,t_*).
}{}
The residue at level $n$ vanishes unless $J(n) < J(n_*)$, since the residues for which $J(n) \geq J(n_*)$ carry vanishing factors of $\alpha(t_*)+r_*=0$.  
Since $\mu(n)$ is without loss of generality strictly monotonic, and we have proven that $\mu(n)\sim J(n)$, it follows that $J(n)$ is strictly monotonic as well, and the sum truncates to the region $n <n_*$.

The above sum rules can be interpreted as a set of equations constraining the residues $R(n,t_*)$, which are taken to be variables.  The total counting is then
\eq{
\# \textrm{ of equations} &= \# \textrm{ of sum rules} =  J(n_*)\\
\# \textrm{ of variables} &=   \# \textrm{ of nonzero residues} = n_*.
}{}
Demanding that the number of constraints not exceed the number of variables, we have 
\be 
J(n_*) \leq n_*. \label{eq:Jleq}
\ee
Since $J(n)$ is monotonic and integer-valued, $J(n+1) - J(n) \geq 1$, which with Eq.~\eqref{eq:Jleq} leaves uniquely
\eq{
J(n) = n.
}{}
This establishes that the spectrum of spins grows linearly with the level.
\medskip

\begin{figure}[t]
    \centering
  \includegraphics[width=0.9\columnwidth]{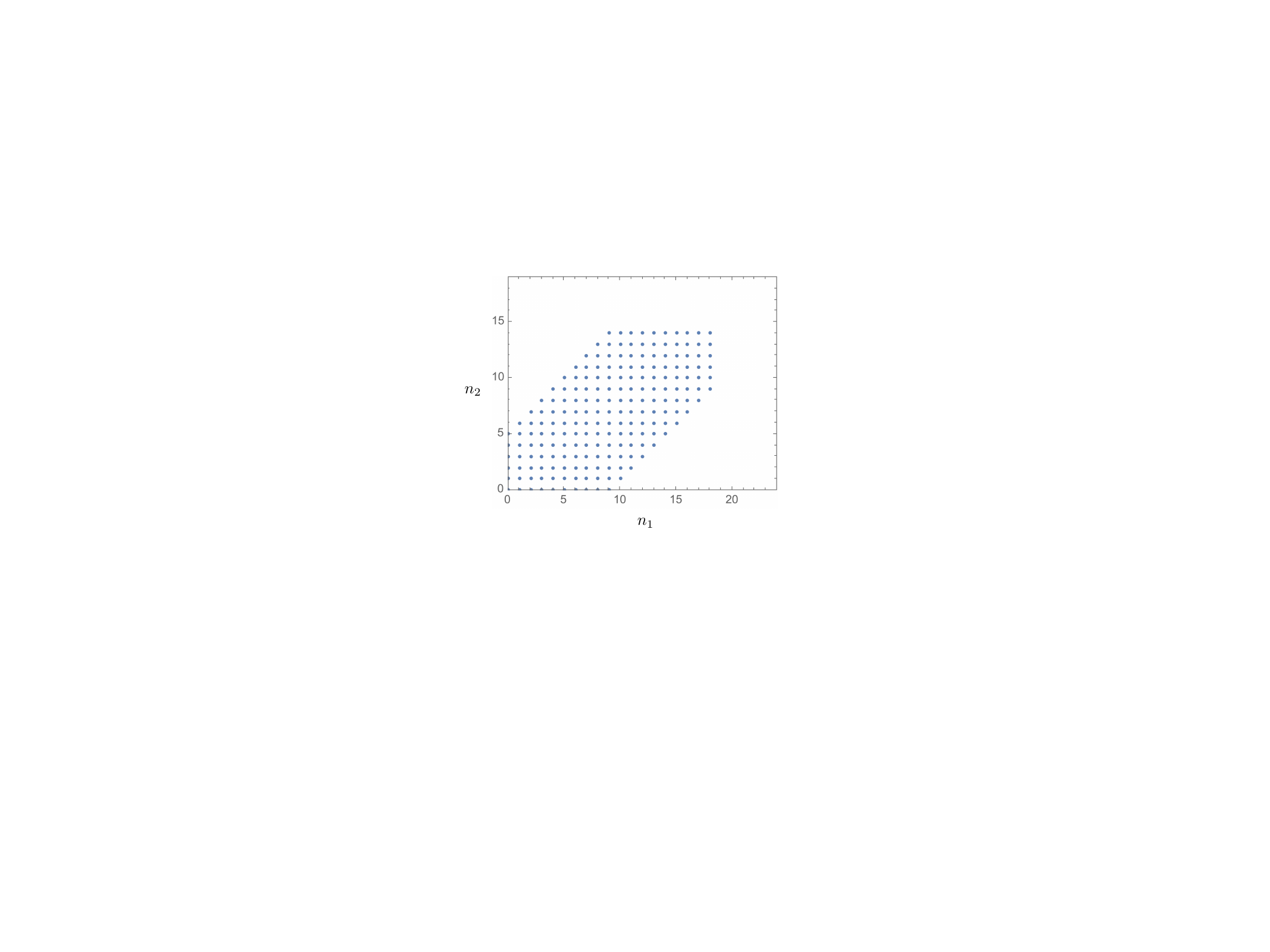} \hspace{10mm}\vspace{-6mm}
    \caption{The domain $D(10,6,25)$, where each dot represents a nonzero residue at level $(n_1,n_2)$ in the kinematic configuration $(t_{23},t_{34},t_{51}) = -(10,6,25)$.  The number of nonzero residues, and hence the dual resonant sum, is finite.
    }
    \label{fig:D}
\end{figure}

\mysec{Appendix B: Five-Point Level Truncation\labeltext{B}{app:5pt_level_truncation}}Let us write the string amplitude as a dual resonant sum~\cite{Arkani-Hamed:2024nzc} of the five-point string residues in \Eq{eq:5ptres}, so
\be 
\begin{aligned}
&A_{\rm str,5}(s_1,s_2,t_{23},t_{34},t_{51}) \\&= \sum_{n_1=0}^\infty \sum_{n_2=0}^\infty \frac{R_{\rm str,5}(n_1,n_2,t_{23},t_{34},t_{51})}{(n_1-s_1)(n_2-s_2)}
\\& = \frac{\Gamma(-s_1)\Gamma(-s_2)\Gamma(-t_{23})\Gamma(-t_{34})}{\Gamma(-s_1-t_{23})\Gamma(-s_2 -t_{34})} \times \\&\qquad \times  {}_3 F_2\left[\begin{array}{c}{\scriptstyle-s_1},{\scriptstyle-s_2},{\scriptstyle-t_{23}-t_{34}+t_{51}} \\ {\scriptstyle-s_1-t_{23}},{\scriptstyle-s_2-t_{34}}\end{array};1\right] .\label{eq:5ptDR}
\end{aligned}
\ee 
The second equality converges when $t_{51} < 0$, but can be extended to arbitrary kinematics via a ${}_6 F_5$ function~\cite{Arkani-Hamed:2024nzc}.
The  five-point residues of the string exhibit a generalization of the level truncation property of four-point amplitudes observed in Ref.~\cite{Cheung:2024uhn}.
In particular, on the kinematic configuration defined by  $(t_{23},t_{34},t_{51}) = -(k_1,k_2,k_3)$ for integers $k_{1,2}\geq 1$ and $k_3\geq k_1+k_2$, all but a finite number of five-point string residues in \Eq{eq:5ptres} vanish.  In particular, the only nonzero residues occur at level $(n_1,n_2)$ in a compact domain $D(k_1,k_2,k_3)$ defined by \eq{
\left\{(n_1,n_2)\left|\begin{array}{c} n_1<k_3-k_2,n_2<k_3-k_1,\\-k_2<n_1-n_2<k_1,\\k_{1,2}\geq 1,k_3 \geq k_1+k_2 \end{array}\right.\right\},
}{eq:D}
depicted in Fig.~\ref{fig:D}.
On these kinematics, the amplitude in Eq.~\eqref{eq:5ptDR} reduces to a rational polynomial in $(s_1,s_2)$.

\medskip

\mysec{Appendix C: Five-Point Normalization\labeltext{C}{sec:5pt_normalizations}}We saw earlier how the five-point Regge zeros are sufficient to uniquely fix the five-point residue of the string amplitude up to a level-dependent normalization factor $c(n_1,n_2)$, so that the five-point bootstrapped amplitude is given by
\be
\begin{aligned} 
&A(s_1,s_2,t_{23},t_{34},t_{51}) =
\\&\qquad \sum_{n_1=0}^\infty \sum_{n_2=0}^\infty \frac{c(n_1,n_2) R_{\rm str,5}(n_1,n_2,t_{23},t_{34},t_{51})}{(n_1-s_1)(n_2-s_2)}.
\end{aligned}
\ee  
Here we will fix these constants as well.

To fix $c(n_1,n_2)$, we exploit the invariance of the amplitude under cyclic permutations of the external legs.  
Level truncation renders the amplitude a rational function of the kinematics.  On this configuration, the constraint of crossing is a solvable algebraic problem.  In particular, 
setting $(s_1,s_2,t_{23},t_{34},t_{51})=-(k_1,k_4,k_2,k_3,k_5)$ for $D(k_2,k_3,k_5)$ and $D(k_3,k_4,k_1)$ defined in Eq.~\eqref{eq:D}, we enforce cyclic invariance of the dual resonant form of the five-point amplitude, $A({-}k_1{,}{-}k_4{,}{-}k_2{,}{-}k_3{,}{-}k_5)= A({-}k_2{,}{-}k_5{,}{-}k_3{,}{-}k_4{,}{-}k_1)$, that is,
\eq{
&&\!\!\!\!\!\!\!\!\!\!\!\! \sum_{\begin{smallmatrix}(n_1,n_2)\in \\ D(k_2,k_3,k_5)\end{smallmatrix}}\frac{c(n_1,n_2)R_{\rm str,5}(n_1,n_2,-k_2,-k_3,-k_5)}{(n_1+k_1)(n_2+k_4)}\phantom{.} \\&= & \!\!\!\!\!\!\!\!\!\!\!\! \sum_{\begin{smallmatrix}(n_1,n_2)\in \\ D(k_3,k_4,k_1)\end{smallmatrix}} \frac{c(n_1,n_2)R_{\rm str,5}(n_1,n_2,-k_3,-k_4,-k_1)}{(n_1+k_2)(n_2+k_5)} .
}{eq:bigcyclic}
Enforcing this condition for all possible choices of integers yields a unique solution, $c(n_1,n_2) = 1$, modulo uniform global normalization.
Hence the five-point string amplitude is uniquely specified by its zeros, together with cyclic symmetry.

\medskip

%%%%%%%%%%%%%%%%%%%%%%%%%%%%%%%%%

\mysec{Appendix D:  $q$-Deformations\labeltext{D}{sec:qdef}}The procedure of bootstrapping the four- and five-point amplitudes from their Regge zeros can be generalized from a linear spectrum to a $q$-spectrum. 
Relaxing bijectivity of $\alpha(t)$ in assumption~\ref{i} and running through our argument in text for the logarithmic Regge trajectory, we have Regge zeros at $\alpha(t)+r = 0$ for integer $r$, whose root scales as $t \propto 1 - q^{-r}$. 
We choose the normalization of $t$ such that these Regge zeros appear at
$t = (1-q^{-r})/(1-q) = [-r]_q$, so the spectrum is given by the $q$-deformed integers, $\mu(n) = [n]_q$, and we have $\alpha(t) =  \log (1+(q-1) t)/\log q$.

Let us first consider four-point scattering.  
Assuming that the residue exhibits only the Regge zeros, $
R(n,t) = c(n) \prod_{r=1}^n (t-[-r]_q)$,
we construct a dual resonant representation of the four-point amplitude, obtaining
\be
\hspace{-2mm}
A(s{,}t) \,{=} \!\sum_{n=0}^\infty \! \frac{R(n,t)}{\mu(n){-}s}  {=}\, q^{\alpha(s)\alpha(t)}\!  \frac{\Gamma_q(\!{-}\alpha(s))\Gamma_q(\!{-}\alpha(t))}{\Gamma_q({-}\alpha(s){-}\alpha(t))}\!,\hspace{-1.5mm}
\label{eq:4ptCoon}
\ee
where the level truncation results of Ref.~\cite{Cheung:2024uhn} have been used with crossing symmetry to uniquely fix $c(n)=q^{n(n+3)/2}/\Gamma_q(1\,{+}\,n)$.  The above expression is precisely the Coon amplitude, which for $q\,{>}\,1$ is dual resonant on kinematics where the sum in Eq.~\eqref{eq:4ptCoon} converges~\cite{Cheung:2024uhn}.

The five-point $q$-deformed amplitude can be written in terms of the $q$-hypergeometric function ${}_3 \phi_2$ as~\cite{Romans,Geiser:2023qqq} \vspace{-1mm}
\be 
\hspace{-1.5mm}\begin{aligned}
&A_{q\textrm{-str,5}}( s_1{,}s_2{,}t_{23}{,}t_{34}{,}t_{51}) \,{=}\vphantom{\big\{\big\}}\, q^{\alpha(\hspace{1pt}\!s_1\!\hspace{1pt})\alpha(\hspace{1pt}\!t_{23}\!\hspace{1pt}){+}\alpha(\hspace{1pt}\!s_2\!\hspace{1pt})\alpha(\hspace{1pt}\!t_{34}\!\hspace{1pt})}\times \\&\vphantom{\Bigg\{\Bigg\}}{\times}\! \frac{\Gamma_q(\!{-}\alpha(s_1))\Gamma_q(\!{-}\alpha(s_2))\Gamma_q(\!{-}\alpha(t_{23}))\Gamma_q(\!{-}\alpha(t_{34}))}{\Gamma_q({-}\alpha(s_1){-}\alpha(t_{23}))\Gamma_q({-}\alpha(s_2){-}\alpha(t_{34}))} {\times} \\&\vphantom{\Bigg\{\Bigg\}} \times \! {}_{3}\phi_{2}\left[\! \begin{array}{c}
\scriptstyle q^{{-}\alpha(s_1)},q^{{-}\alpha(s_2)},q^{\alpha(t_{51}){-}\alpha(t_{23}){-}\alpha(t_{34})}\\
\scriptstyle q^{{-}\alpha(s_1){-}\alpha(t_{23})},q^{{-}\alpha(s_2){-}\alpha(t_{34})}
\end{array}\!\!;q;q\right].
\end{aligned}\hspace{-1.5mm}
\ee
For $q>1$, this amplitude is dual resonant for kinematics where the sum converges, equaling
\be
\sum_{n_1{=}0}^\infty \sum_{n_2{=}0}^\infty \frac{R_{q\textrm{-str,5}}(n_1,n_2,t_{23},t_{34},t_{51})}{([n_1]_q {-} s_1)([n_2]_q {-} s_2)},\vspace{1mm}
\ee
for the corresponding five-point residue polynomial,\vspace{-1mm}
\be
\begin{aligned}
&R_{q\textrm{-str,5}}(n_1,n_2,t_{23},t_{34},t_{51}) =\vphantom{\bigg\{\bigg\}}  \\& \vphantom{\Bigg\{\Bigg\}}\frac{(-1)^{n_1+n_2}q^{\frac{n_1(n_1+3)+n_2(n_2+3)}{2}+n_1\alpha(t_{23})+n_2\alpha(t_{34})}}{\Gamma_q(n_1+1)\Gamma_q(n_2+1)} \times \\& \vphantom{\Bigg\{\Bigg\}}\times\frac{\Gamma_{q}(-\alpha(t_{23}))\Gamma_{q}(-\alpha(t_{34}))}{\Gamma_{q}(-n_1-\alpha(t_{23}))\Gamma_{q}(-n_2-\alpha(t_{34}))} \times \\&\vphantom{\Bigg\{\Bigg\}}\times {}_{3}\phi_{2}\left[\begin{array}{c}
\scriptstyle q^{{-}n_1},q^{{-}n_2},q^{\alpha(t_{51}){-}\alpha(t_{23}){-}\alpha(t_{34})}\\ \scriptstyle
q^{{-}n_1{-}\alpha(t_{23})},q^{{-}n_2{-}\alpha(t_{34})}
\end{array};q;q\right].
\end{aligned}\label{eq:RC5pt}
\ee
Simply replacing $t_{ij}$ with $\alpha(t_{ij})$ in the definitions of $a_1,a_2,a_{12}$, so $a_1= \alpha(t_{51}) -\alpha(t_{34})$, $a_2= \alpha(t_{51}) - \alpha(t_{23})$, and $a_{12} =\alpha(t_{23}) + \alpha(t_{34}) - \alpha(t_{51})$, we find that this residue exhibits precisely the same Regge zeros as the five-point string amplitude, namely, $a_1+r_1=a_2+r_2=a_{12}+r_{12}-r_1-r_2=0$, that is, $t_{23} = [r_2-r_{12}]_q$, $t_{34} = [r_1-r_{12}]_q$, and $t_{51} = [-r_{12}]_q$, all for integers $|r_1|\leq n_1$, $|r_2|\leq n_2$, and $1\leq r_{12}\leq r_1+r_2$.

Remarkably, we find that the $q$-deformed five-point amplitude can be constructed uniquely from these zeros, just like the string. Again taking the ansatz residue $R(n_1,n_2,t_{23}, t_{34}, t_{51})= \sum_{i=0}^{n_1} \sum_{j=0}^{n_2} \sum_{k=0}^{{\rm min}(n_1-i,n_2-j)} \lambda_{i,j,k}(n_1,n_2) t_{23}^i t_{34}^j t_{51}^k$ for  states of spin up to $n_{1,2}$ on the factorization channel at $s_{1,2} = [n_{1,2}]_q$, we find that the unique solution satisfying the zeros described above is the residue in \Eq{eq:RC5pt}, up to some unfixed normalization $c(n_1,n_2)$.

Furthermore, on the special kinematic configuration $(t_{23},t_{34},t_{51}) = ([-k_1]_q,[-k_2]_q,[-k_3]_q)$, the only nonzero residues are precisely those defined in $D(k_1,k_2,k_3)$ in Eq.~\eqref{eq:D}. That is, these $q$-deformed five-point residues exhibit level truncation, so as before we can algebraically fix the $c(n_1,n_2)$ normalizations by demanding cyclic invariance.  Here we again employ \Eq{eq:bigcyclic}, albeit with the residues in Eq.~\eqref{eq:RC5pt} and with the denominators replaced by factors of $[n]_q - [-k]_q$.
This yields a unique solution consistent with cyclic invariance for which $c(n_1,n_2) = 1$, modulo global normalization. Thus, the $q$-deformed five-point amplitude is uniquely constructible from its Regge zeros and cyclic symmetry.

\end{document}